# Surfaces of colloidal PbSe nanocrystals probed by thin-film positron annihilation spectroscopy

L. Chai[1], W. Al-Sawai[2], Y. Gao[3,4], A. J. Houtepen[3], P. E. Mijnarends[1,2], B. Barbiellini[2], H. Schut[1], L. C. van Schaarenburg[1], M. A. van Huis[4], L. Ravelli[5], W. Egger[5], S. Kaprzyk[6,2], A. Bansil[2], and S. W. H. Eijt[1,a)]

[1]*Department of Radiation, Radionuclides & Reactors, Faculty of Applied Sciences, Delft University of Technology, Mekelweg 15, NL-2629 JB, Delft, The Netherlands*

[2]*Physics Department, Northeastern University, Boston, Massachusetts, 02115, USA*

[3]*Department of Chemical Engineering, Faculty of Applied Sciences, Delft University of Technology, Julianalaan 136, NL-2628 BL, Delft, The Netherlands*

[4]*Kavli Institute of Nanoscience, Faculty of Applied Sciences, Delft University of Technology, Lorentzweg 1, NL-2628 CJ, Delft, The Netherlands*

[5]*Institut für Angewandte Physik und Messtechnik, Universität der Bundeswehr München, Werner-Heisenberg-Weg 39, D-85579, Neubiberg, Germany*

[6]*Academy of Mining and Metallurgy AGH, PL-30059, Kraków, Poland*

---

a) Author to whom correspondence should be addressed. Electronic mail: S.W.H.Eijt@tudelft.nl

Positron annihilation lifetime spectroscopy (PALS) and positron-electron momentum density (PEMD) studies on multilayers of PbSe nanocrystals (NCs), supported by transmission electron microscopy (TEM), show that positrons are strongly trapped at NC surfaces, where they provide insight into the surface composition and electronic structure of PbSe NCs. Our analysis indicates abundant annihilation of positrons with Se electrons at the NC surfaces and with O electrons of the oleic ligands bound to Pb ad-atoms at the NC surfaces, which demonstrates that positrons can be used as a sensitive probe to investigate the surface physics and chemistry of nanocrystals inside multilayers. Ab-initio electronic structure calculations provide detailed insight in the valence and semi-core electron contributions to the positron-electron momentum density of PbSe. Both lifetime and PEMD are found to correlate with changes in the particle morphology characteristic of partial ligand removal.

## I. INTRODUCTION

Nanocrystals (NCs) of inorganic compound semiconductors are of considerable interest to the development of advanced solar cells since their tunable optical properties and relatively low production cost can potentially boost the efficiency and lower the price of solar cells.[1-5] Colloidal PbSe nanocrystals are promising for solar cell applications, since they show a band gap that is largely tunable in the range from 0.28 eV up to at least 1.2 eV (depending on particle size), which covers an important part of the solar spectrum.[5-8] Moreover, PbSe NCs have demonstrated carrier multiplication, i.e., the generation of more than one electron-hole pair per single absorbed photon.[9,10] The electro-optical properties of colloidal semiconductor NCs are in general largely affected by the surface structure and composition including the passivation of surface states by ligand molecules attached to their surfaces. These surface properties depend, in turn, on sample preparation.[4,11-13] For example, the morphology and surface composition of PbSe or CdSe NCs depend strongly on the method of synthesis and capping material used.[5,11,12,14] Several groups have found non-stoichiometric ratios of Pb to Se atoms in PbSe NCs in solution, which leads to the hypothesis that the NCs are enclosed by a monolayer of Pb atoms.[15-18] On the other hand, synchrotron X-ray Photoelectron Spectroscopy studies on PbSe NC layers have indicated the formation of a Pb-deficient sub-surface layer terminated by Se-atoms at the surface.[19]

Recent experimental studies indicate that the positron is a sensitive probe of the electronic structure of colloidal NCs[20-22] and of the chemical composition at their surfaces, based on preferential trapping and annihilation of positrons at the surfaces.[21-23] In this paper we show that positrons are strongly trapped at the surfaces of the PbSe NCs, where they annihilate mostly with the Se atoms and with the O atoms from the oleate (OA) ligands bound to Pb ad-atoms. The dominant trend in the variation in positron annihilation characteristics is induced by partial removal of oleate ligands together with the attached Pb ad-atoms, leaving Se-rich surfaces behind, which is consistent with the changes in morphology of the PbSe NCs observed by Transmission Electron Microscopy (TEM).

## II. EXPERIMENT and CALCULATION

Samples of PbSe NCs with four different average sizes ranging between 2.8 nm and 9.7 nm were synthesized using the method of Talapin and Murray.[24] A mixture of 95% hexane and 5% octane was used as a solvent for drop-casting PbSe NCs onto indium-tin-oxide (ITO) coated glass substrates.[25] Films were formed with thicknesses in the range from 200 nm to over 1000 nm, as determined from positron Doppler broadening depth-profiles. The particle sizes were determined by Optical Absorption Spectroscopy, X-ray Diffraction (XRD) and TEM using a FEI Titan high-resolution TEM at 300 kV. The drop-cast nanocrystal films were further examined using positron annihilation lifetime spectroscopy (PALS) on the pulsed low-energy positron lifetime spectrometer PLEPS at FRM-II in Garching and two-dimensional angular correlation of annihilation radiation (2D-ACAR) on the thin-film POSH-ACAR setup at the Reactor Institute Delft. For comparison, a PbSe single crystal was studied by 2D-ACAR using a $^{22}$Na positron source. The positron lifetime spectra were fitted using the program POSWIN.[25-27] The 2D-ACAR data were analysed using the program ACAR2D.[28] Ab-initio calculations of the PEMD were performed using the Korringa-Kohn-Rostoker method[29,30] to extract robust electronic contributions to the PEMD of the PbSe NCs. To further understand the nature of divacancies in PbSe, we employed the ab initio method as implemented in the VASP code[31] supercell calculations of Schottky pair defects ($V_{Pb}V_{Se}$) in PbSe.[25]

## III. RESULTS and DISCUSSION

The PALS spectra of nearly all thin layers of PbSe NCs were found to display a dominant positron lifetime component $\tau_2$ with a lifetime of 340 to 380 ps using a three component lifetime analysis. Figure 1 presents the average lifetime $<\tau> = (I_1\tau_1 + I_2\tau_2)/(I_1 + I_2)$ of the two short lifetime components, i.e. excluding the long lifetime ortho-Ps (o-Ps) component $\tau_3$, as a function of size of the nanocrystals. The weighted average short lifetime $<\tau>$ is a statistically accurate parameter. The high intensity $I_2$ (85% to 99%) of the dominant lifetime component $\tau_2$ indicates near-saturation trapping of positrons.[25] We compared the observed intensities to estimated fractions of positrons stopped in the PbSe core extracted from the mass-density-weighted volume fractions of the PbSe NC cores and the OA shells, respectively.[25] Indeed, the estimated fractions provide a good description of the intensity of the lifetime component $\tau_2$. This is a first indication that the positrons predominantly annihilate with electrons of the PbSe nanocrystals.

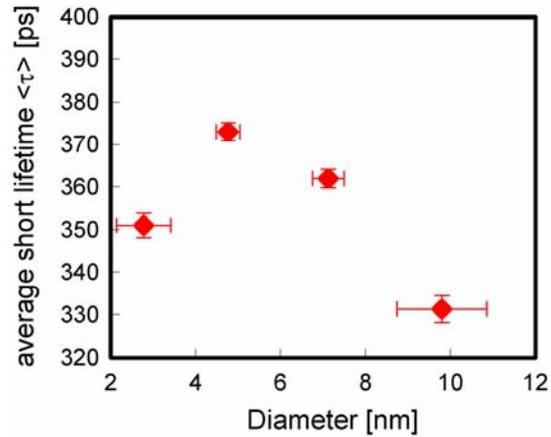

FIG. 1. Average of the short positron lifetimes $<\tau>$ (◆) as a function of nanocrystal diameter.

Our positron lifetime analysis further revealed the presence of an o-Ps lifetime component with intensity $I_3$ in the range of 1.4% to 4.4% and a lifetime $\tau_3$ in the range of 2.0 ns to 2.7 ns. Positronium is formed naturally in the open spaces available between the carbon chains of the OA ligands. Studies on fatty acids of a comparable chain length[32] indeed revealed significant o-Ps formation with lifetimes typically in the range of 1.5-2 ns. The relatively modest o-Ps intensities observed here agree with the range of intensities expected from the estimated fractions of positrons stopped in the OA ligands.

The observed lifetimes $\tau_2$ associated with positrons stopped in the PbSe cores are much longer than the calculated lifetimes reported for bulk mono-crystalline PbSe (213 ps) and for Pb and Se mono-vacancies in PbSe (298 ps and 272 ps, respectively[33]). Our calculations based on the Generalized Gradient Approximation[34,35] revealed a positron lifetime for bulk PbSe of 223 ps, in agreement with experiment.[33] The high values for the positron lifetime observed in this study show that annihilation takes place in open spaces that are equal to at least the size of a di-vacancy[33,36] ($V_{Pb}V_{Se}$) and act as trapping centers for the positrons stopped inside the PbSe NC. This strongly indicates that the sites at which the positrons annihilate are located predominantly at the surfaces of the PbSe nanocrystals. An alternative explanation would consist of annihilation in vacancy clusters inside the PbSe NCs. It is, however, highly improbable that formation of the latter results from this type of synthesis. Considering the small diffusion distances required, vacancies, when generated during the growth of the NCs, would quickly diffuse to their surfaces at the synthesis temperatures used (between 100°C and 165°C). Moreover, since the trapping is saturated, the second explanation would mean that nearly all PbSe NCs are highly defective. A simple estimate shows that an unphysically large divacancy concentration of the order of one divacancy per 100 Pb atoms (for the smallest NCs) would be needed to explain the size of the observed effect, i.e. at least two orders of magnitude larger than Pb *mono*vacancy concentrations observed in PbSe single crystals.[33] Indeed, the formation energy of a $V_{Pb}V_{Se}$ divacancy of 1.47 eV was found from our VASP calculation.[25] This implies that even at the highest synthesis temperature, the equilibrium concentration of divacancies is less than $10^{-16}$ per pair of Pb-Se atoms, excluding the divacancy scenario. Therefore, the positron lifetime results demonstrate that positrons annihilate predominantly at the surfaces of the PbSe NCs and are not confined inside the PbSe quantum dots. This validates previous hypotheses based on positron studies of the PEMD of CdSe NCs, and provides a solid basis for the use of positron annihilation as a sensitive tool for probing surface properties of semiconductor nanoparticles.

In order to gain additional insight in the surface composition of the PbSe nanocrystals, we investigated the same set of films using the positron 2D-ACAR method. The measured 2D-ACAR momentum distributions were found to be isotropic, consistent with the random orientation of the NCs obtained by XRD. In Figure 2, the evolution of the 1D-ACAR momentum distributions $N(p)$ as a function of particle size of the PbSe NCs is presented as the ratio between $N(p)$ and the directionally averaged 1D-ACAR momentum distribution for the PbSe single crystal. At low momenta, the ratio curves show a reduction in the momentum range below 0.6 a.u. and a corresponding peak at ~1.0 a.u. These features are characteristic of broadening of the PbSe momentum distribution, primarily caused by quantum confinement of the Se(4p) valence electron orbitals (similar to CdSe NCs[21,22]), and corresponding band-edge blurring, i.e., momentum density smearing at the boundary of the Jones zone.[20-22] The area of the peak at $p$~1.0 a.u.[25] follows quite well a $1/d$ dependence on particle diameter $d$, as could be expected from the scaling law for the optical band gap $\Delta E_{gap}$~$1/d$ for PbSe nanoparticles.[5]

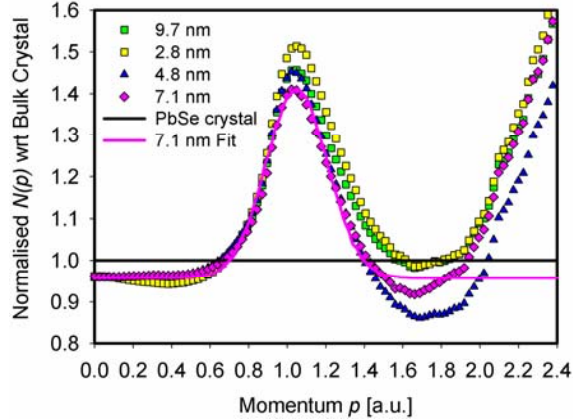

FIG. 2. 1D-ACAR momentum distributions $N(p)$ of PbSe nanocrystals presented as ratio to the 1D-ACAR distribution of a bulk PbSe single crystal measured with a [22]Na source. The confinement peak near $p$~1.0 a.u. is fitted by a Gaussian curve to extract the peak area as a function of particle size.

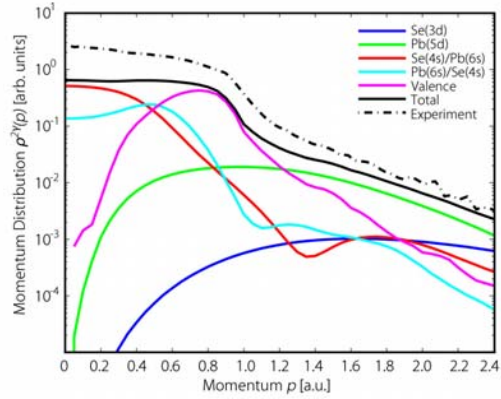

FIG. 3. Directionally averaged positron-electron momentum distribution $\rho^{2\gamma}(p)$ of bulk crystalline PbSe from first principles calculations (heavy black solid line) and from a 2D-ACAR experiment using a [22]Na source (heavy dash-dotted line). The experimental curve has been shifted vertically for clarity. Colored curves give the calculated contributions from the upper valence orbitals and from the Pb(6s), Pb(5d), Se(4s) and Se(3d) orbitals. The valence band contribution is dominated by contributions of the Se(4p) and, to a lesser extent, Pb(4p) electron orbitals.

Figure 3 shows the calculated, directionally averaged, positron-electron momentum density $\rho^{2\gamma}(p)$ for the PbSe single crystal, broken down into the contributions of the various occupied electron bands. The Pb(5d) and Se(3d) core orbitals are much less sensitive to the presence of a surface than the valence electron orbitals, which are modified by the different atomic bonds formed at the surfaces. Their contributions to the observed PEMD therefore provide a good measure for the local chemical composition at the annihilation site. The calculations indicate that about 80% of the momentum distribution $\rho^{2\gamma}(p)$ in the range from 1.8 to 2.3 a.u. stems from electrons in Pb(5d) orbitals. We use this interval to quantify the contribution of Pb atoms to the surface composition of the PbSe NCs as seen by positron annihilation, using the parameter $I_{Pb}$ defined as follows:[21]

$$I_{Pb} = \int_{1.8\ a.u.}^{2.3\ a.u.} \rho^{2\gamma}(p) p^2 dp \tag{1}$$

in which the isotropic momentum distribution $\rho^{2\gamma}(p) = -\frac{1}{p_z}\frac{dN(p_z)}{dp_z}\bigg|_{p_z=p}$ of the PbSe NCs is extracted from the measured 1D-ACAR momentum distribution $N(p_z)$.[37]

Figure 4 shows that $I_{Pb}$ for the PbSe NCs is significantly smaller (~25%) than that for bulk PbSe and shows a significant variation with diameter. The observed reduction in $I_{Pb}$ indicates an increase in positron annihilation near Se-atoms instead of Pb-atoms at the surfaces of the PbSe NCs. We note here that surface relaxations, which were the primary cause for the – stronger and more constant – reduction in $I_{Cd}$ for CdSe NCs,[21,22] are less pronounced for PbSe NCs, and, as illustrated by Franceschetti in Ref. 7, are different from those of CdSe NCs. The self-healing mechanism producing a Se rich monolayer at the surface of the CdSe NCs does not apply to the PbSe NCs. Here, the 20-35 % reduction of Pb signal is naturally explained in terms of a competition between OA ligand coverage and Se exposed areas at the NC surface as will be explained below. Interestingly, in the CdSe case, the ligand effects were much smaller.[38]

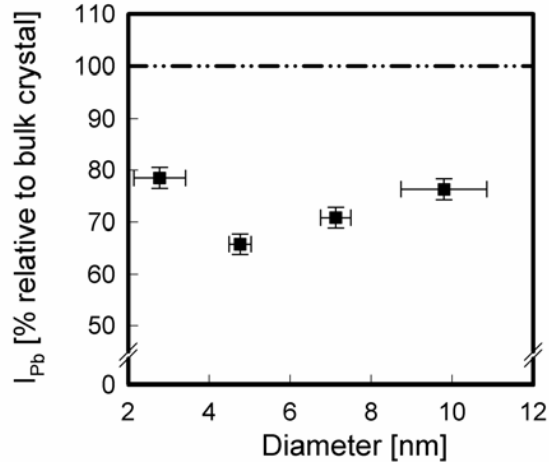

FIG. 4. Contribution of the Pb(5d) electron orbitals to the ACAR momentum distribution as a function of nanocrystal size relative to that for the PbSe bulk single crystal (dash-dotted line), extracted from the intensities of the momentum distribution $\rho^{2\gamma}(p)$ in the range from 1.8 to 2.3 a.u.

In Fig. 4, the Pb contribution for the 4.8 nm PbSe NCs is most strongly reduced, which points to a partial decapping of the OA ligands from the PbSe NCs. Recent advanced characterization of quantum dots (QDs)[15-17] showed that a PbSe nanoparticle in solution is composed of a PbSe core and an individual surface layer of Pb atoms bonded to the oxygen atom of the OA ligand, i.e. the OA ligands are (most likely) bound to the nanocrystal surface in the form of lead oleate.[39,40] The ratio curves presented in Figure 2 thus indicate that the positron annihilation at the surface occurs predominantly at Se and O sites, involving (1) annihilation at Se-rich areas at the surface of PbSe nanocrystals, and (2) annihilation at oxygen atoms of the OA molecules on top of Pb ad-atoms at specific PbSe nanocrystal facets which are well-covered with OA ligands. The latter is inferred from the steep increase at high momenta (p>2 a.u.) in the ratio curves of Figure 2, which is a signature of positrons annihilating with the core electrons of oxygen.[41] Indeed, ab-initio calculations of the PEMD of atomic oxygen[25] reveal a clear rise above p>2 a.u. in the ratio curve with respect to bulk PbSe annihilations. This is remarkably similar to the observed high momenta increase in Figure 2, induced by the contribution of the O(1s) core states to the observed PEMD of the OA-capped PbSe nanocrystals. We note that the observation of the O(1s) contribution in the ratio curves again points to positron trapping at the surfaces of the OA-capped PbSe nanocrystals, since such a signature would be evidently absent for positron annihilation in $V_{Pb}V_{Se}$ divacancies inside the PbSe NCs.

Interestingly, the dependence of $I_{Pb}$ on particle size is seen to correlate with that of the positron lifetime.[25] A clear trend of an increasing positron lifetime (Figure 1) and a decreasing Pb contribution (Figure 4) is observed in the particle size range from 9.7 nm down to 4.8 nm, while the reverse trend occurs when approaching the smallest particle size. From this we infer that, while positrons trap at sites with increasing amounts of surrounding Se-atoms, the annihilation rate simultaneously decreases towards that of the 4.8 nm sample, a sign of the development of more open space, which is most probably induced by removal of some of the ligand molecules. Such a partial removal may result from the washing of PbSe nanocrystals, or from the drop-cast deposition and subsequent drying of the PbSe nanocrystal film on the ITO coated substrate.[25] As the surface Pb atoms are bound with equal strength to the oleate ligands and the neighboring Se atoms, the washing or deposition may result in the dissociation of ligands from the surface Pb atoms in parallel with the dissociation of Pb-ligand units from the QD, leaving a primarily Se-terminated nanocrystal surface behind. A similar removal of Pb from the surface has also been reported in PbS QDs by various authors.[42,43] Considering the highly ionic character of Pb-Se bonds in PbSe, Se-rich and incompletely capped nanocrystal facets will attract positrons.

Moreover, the observed trends also correlate with the variation in morphology of the rock salt PbSe NCs seen with TEM. The 9.7 nm NCs are typically well-separated and have a nearly spherical multifaceted morphology.[25] High-resolution images[25] for the 9.7 nm sample show a multifaceted morphology of the nanocrystal which is very similar to those of well-capped PbSe nanocrystals in Ref. 14. The latter study inferred that a good ligand capping of PbSe NCs results in nearly equal surface energies for the {100}, {110} and {111}-facets, leading to the observed multifaceted morphology of the PbSe NCs.[14] The 7.1 nm particles, in contrast, are multifaceted truncated octahedrons with pronounced {111}-facets and small {100}-facets. This is typical for a reduced oleic acid ligand coverage of the {100}-surfaces, as described in detail by Bealing *et al*.[39] The 4.8 nm particles show more variations in shape and are less centrosymmetric than the larger particles. In particular, well-developed PbSe {110}-edges can be discerned. The presence of different types of partially uncovered facets will alter the annihilation characteristics. For example, a pair-wise grouping of Se-atoms exists in the PbSe {110}-surface,[14] which is expected to lead to enhanced overlap of the positron wave function with Se. The variation in positron lifetime and $I_{Pb}$ can thus be explained by the development of different types of facets with an attractive Se-termination, induced by a partial removal of OA ligands. In particular, the 4.8 nm sample (with clear asymmetric particle shapes[25]) shows a very high fraction of 98% of the positrons annihilating at the surface and the lowest o-Ps fraction $I_3$ typical for annihilation with the tails of the OA ligands. It also shows the highest average short positron lifetime <τ> and a very low Pb-contribution $I_{Pb}$. These features are indicative of reduced ligand capping and annihilation at Se-rich facets. Indeed, the increase in the ratio curve at high momenta in the range of ~2.2 a.u. and beyond (Figure 2) is consistently the weakest for the 4.8 nm sample, which constitutes a clear signature for reduced annihilation at O atoms.

**CONCLUSION**

In summary, we used positron lifetime spectroscopy to demonstrate that positrons trap at the surface of PbSe NCs, where they provide insight into the surface composition and electronic structure of PbSe NCs. We found a reduction in annihilations from Pb(5d) electrons of the NCs, indicating that positrons annihilate mostly with Se electrons in the surface layer and with electrons of O atoms from the oleate ligands bound to Pb ad-atoms at the NC surfaces. A clear correlation was found between the positron lifetime, the Pb(5d) contribution and the morphology of the PbSe NCs, indicating that the formation of facets leads to a larger local open volume at Se-rich annihilation sites. This is induced by partial removal of oleate ligands together with the attached Pb ad-atoms, leaving Se-rich surfaces behind. The present study reveals that positron annihilation can be used as an advanced characterization tool to unravel many novel properties associated with the surface physics and chemistry of nanocrystals inside multilayers. Ab-initio methods are currently under development in order to achieve a refined knowledge of the positron wave function and its overlap with the electron orbitals of surface atoms, paving the way to extract and monitor accurate quantitative surface compositions of colloidal nanocrystals in thin films.

## ACKNOWLEDGMENTS

This work is supported by the US Department of Energy, Office of Science, Basic Energy Sciences contract DE-FG02-07ER46352. It benefited from Northeastern University's Advanced Scientific Computation Center (ASCC), theory support at the Advanced Light Source, Berkeley, and the allocation of supercomputer time at NERSC through grant number DE-AC02-05CH11231, and at Stichting Nationale Computerfaciliteiten (National Computing Facilities Foundation NCF) with financial support from the Netherlands Organization for Scientific Research (NWO). This work was supported by the Polish National Science Center (NCN) under the grant DEC-2011/02/A/ST3/00124. We acknowledge Dr. Gleb Gribakin for providing us with the calculated O(1s), O(2s) and O(2p) positron-electron momentum densities (PEMDs) of atomic oxygen.

**FIGURE LEGENDS**

**FIG. 1.** (a) Average of the short positron lifetimes $<\tau>$ (◆) as a function of nanocrystal diameter.

**FIG. 2.** 1D-ACAR momentum distributions $N(p)$ of PbSe nanocrystals presented as ratio to the 1D-ACAR distribution of a bulk PbSe single crystal measured with a $^{22}$Na source. The confinement peak near $p\sim 1.0$ a.u. is fitted by a Gaussian curve to extract the peak area as a function of particle size.

**FIG. 3.** Directionally averaged positron-electron momentum distribution $\rho^{2\gamma}(p)$ of bulk crystalline PbSe from first principles calculations (heavy black solid line) and from a 2D-ACAR experiment using a $^{22}$Na source (heavy dash-dotted line). The experimental curve has been shifted vertically for clarity. Colored curves give the calculated contributions from the upper valence orbitals and from the Pb(6s), Pb(5d), Se(4s) and Se(3d) orbitals. The valence band contribution is dominated by contributions of the Se(4p) and, to a lesser extent, Pb(4p) electron orbitals.

**FIG. 4.** Contribution of the Pb(5d) electron orbitals to the ACAR momentum distribution as a function of nanocrystal size relative to that for the PbSe bulk single crystal (dash-dotted line), extracted from the intensities of the momentum distribution $\rho^{2\gamma}(p)$ in the range from 1.8 to 2.3 a.u.

# SUPPLEMENTARY MATERIAL

*Surfaces of colloidal PbSe nanocrystals probed by thin-film positron annihilation spectroscopy*, L. Chai *et al*.

## A. Synthesis and characterization

Samples of PbSe NCs with four different average sizes ranging between 2.8 nm and 9.7 nm were synthesized using the method of Talapin and Murray.[1] The crude solution from the synthesis was washed two to four times using butanol and methanol to remove excess ligands. Particles were collected by centrifugation. A mixture of 95% hexane and 5% octane was used as a solvent for drop-casting NCs onto an indium-tin-oxide (ITO) coated glass substrate. The PbSe nanocrystal films thus formed were left in a closed Petri dish for drying by normal solvent evaporation. An additional sample was synthesized with a fifth average particle size of 4.1 nm, which was dried under vacuum or by heating to 100°C for about three hours due to difficulties encountered in drying by normal means.

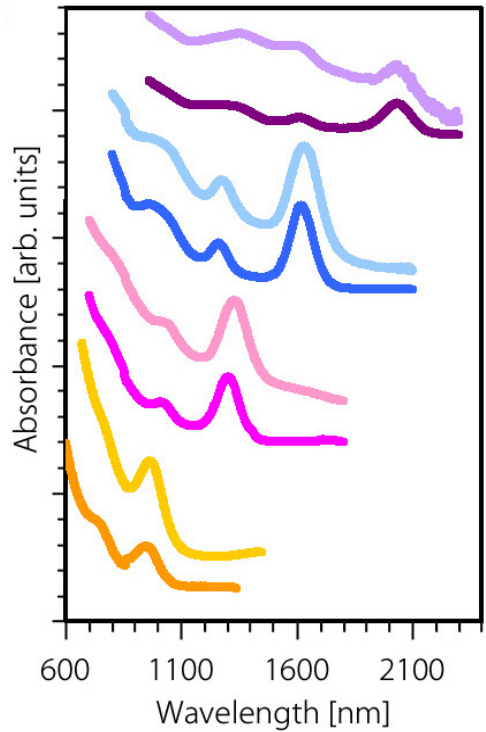

**Figure S1.** Pairs of optical absorption spectra of PbSe NCs in solution (lower curve of each pair) and in thin-films (upper curve of each pair), offset and scaled vertically. From top to bottom the NC sizes are 7.1 nm (plum), 4.8 nm (blue), 4.1 nm (pink), and 2.8 nm (gold).

Utmost care was taken to prevent exposure of the samples to air, which was limited to the short periods during which the samples were handled while setting up the 2D-ACAR and Doppler broadening experiments and during collection of the optical absorption spectroscopy (OAS) spectra.

An extract of each of the nanocrystal solutions was measured after the first wash cycle using OAS as a conventional method of determining the sizes of PbSe NCs. The optical absorption spectra were recorded with a Perkin-Elmer Lambda 900 spectrometer equipped with an integrating sphere. The PbSe nanocrystal films formed by drop-casting were analyzed using XRD and depth-profiling positron spectroscopy.[2,3] The films had thicknesses in the range from 200 nm to over 1000 nm, as determined from positron Doppler broadening depth-profiles. OAS spectra were collected twice, the first time on nanocrystal solutions from synthesis and the second time on the drop-cast thin films after all PAS measurements were performed, *i.e.* prior to re-dissolving the films using hexane for transferral onto grids for the final TEM experiments.

TEM images were collected using a FEI Titan high-resolution transmission electron microscope at 300 kV. Diameters of individual particles were measured using the MATLAB program HREMIP, which computes and stores the distances between consecutive mouse clicks on a TEM image. Size information could only be obtained for the three larger particle sizes due to difficulties in capturing high-contrast images of small particles. Taking the average of the OAS and TEM results, the largest three particle sizes were determined to be 9.7 ± 1.1, 7.1 ± 0.4 and 4.8 ± 0.3 nm. The two smaller sizes were derived solely from the OAS results, namely 4.1 ± 0.3 and 2.8 ± 0.6 nm.

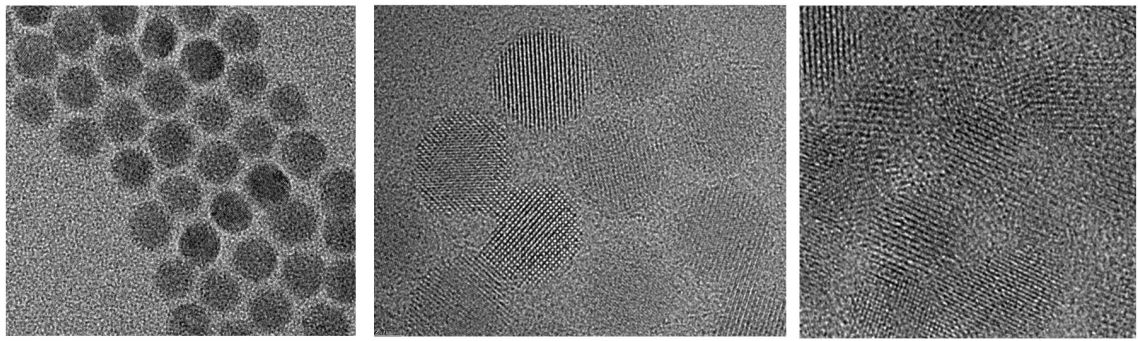

**Figure S2.** TEM images of samples with 9.7 nm (left), 7.1 nm (centre) and 4.8 nm (right) PbSe nanocrystals.

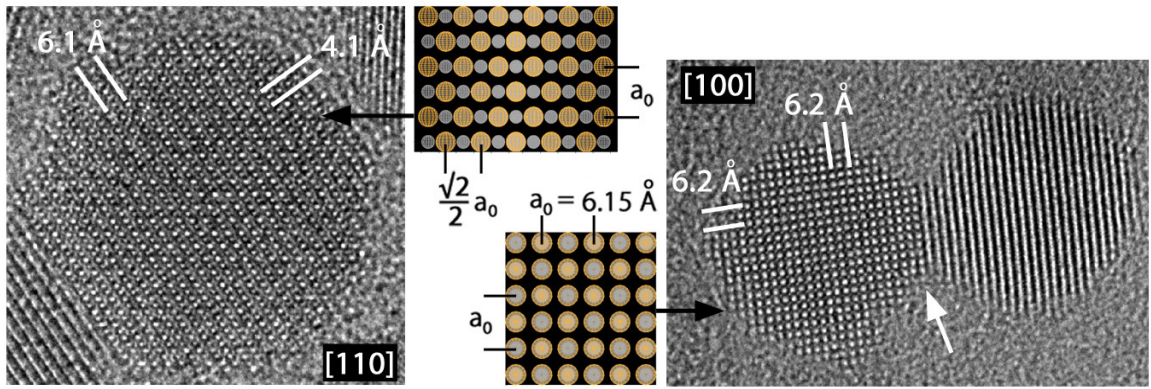

**Figure S3.** High-resolution images of PbSe nanocrystals with [110] and [100] orientations for the 9.7 nm and 7.1 nm sample respectively. Lattice spacings match those of rock-salt PbSe.

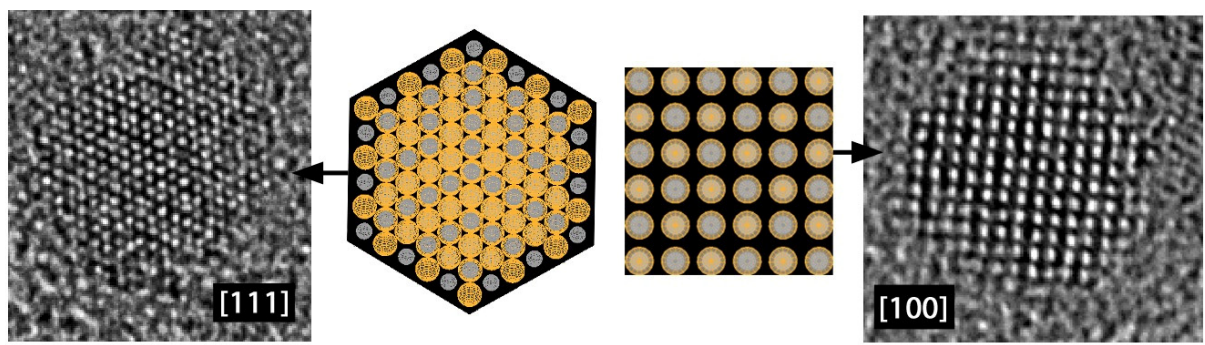

**Figure S4.** High-resolution images of PbSe nanocrystals with [111] and [100] orientations for the 4.8 nm sample, respectively. Lattice spacings match those of rock-salt PbSe.

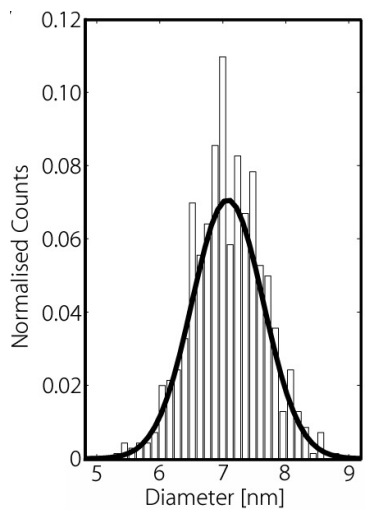

**Figure S5.** Size distribution extracted from 702 nanocrystals in TEM images of the 7.1 nm sample. A mean size of 7.1 nm and a standard deviation of 0.6 nm are obtained from the Gaussian fit.

**B. Positron annihilation lifetime spectroscopy**

The drop-cast PbSe nanocrystal films were examined using positron annihilation lifetime spectroscopy (PALS). In a positron lifetime experiment, a population of $N(0)$ positrons implanted inside a medium decays exponentially according to

$$-dN(t)/dt = N(0)\sum_{i}(I_i/\tau_i)\exp(-t/\tau_i), \quad \sum_{i} I_i = 1,$$

where $I_i$ is the fraction of positrons decaying with lifetime $\tau_i$. A sample can have multiple types of annihilation sites, each giving rise to a different lifetime $\tau_i$ and corresponding fraction $I_i$. In a simplified model $\tau_i$ at site $i$ is inversely proportional to the electron density at that site. Large vacancies and open volumes have lower electron densities and thus a longer lifetime. PALS measures these lifetimes.[4]

The lifetimes in the PbSe thin-film samples were measured using the pulsed low energy positron lifetime spectrometer PLEPS[5] coupled to the high-intensity slow positron beam NEPOMUC at the reactor FRM-II in München with positron implantation energies ranging from 0.5 to 18 keV. We studied a set of nine PbSe nanocrystal samples with a size of 9×9 mm². The average sizes of the NCs ranged from 2.8 nm to 9.7 nm. Reference spectra were obtained on both p-SiC and metglass samples at several positron implantation energies to determine the instrumental time resolution function.[6] The raw data were fitted using the program POSWIN[7] to obtain the number of lifetime components and their respective lifetimes and intensities.[6] Each implantation energy produced one time-resolved spectrum. Each spectrum analysis resulted in (at most) three lifetime components.

Table S1 presents the derived positron lifetimes and intensities extracted for each particle size from the fitted values of lifetime plateau in the PALS depth profiles which corresponds to the QD layer, based on averaged values for one or two samples. The high intensity $I_2$ (85% to 99%) of the dominant lifetime component $\tau_2$ indicates near-saturation trapping of positrons at the surfaces of the PbSe nanocrystals. Further, an o-Ps lifetime component with intensity $I_3$ in the range of 1.4% to 4.4% and a lifetime $\tau_3$ in the range of 2.0 ns to 2.7 ns was present for all samples. In the 9.7 nm PbSe nanocrystal sample, we also observed a small short lifetime component with a positron lifetime close to that of defect-free bulk PbSe. This indicates that some bulk-like annihilation takes place for the largest PbSe nanocrystals. Finally, in the lifetime spectra for the 7.1 nm PbSe nanocrystals a lifetime component with $\tau_2 \sim 550$ ps was present characteristic for the statistical average of o-Ps and p-Ps annihilation.

**Table S1**: Positron lifetimes and intensity of lifetime components for PbSe nanoparticles (analysis using three lifetime components)

| Particle diameter | $\tau_2$ [ps] | $I_2$ (%) | $\tau_3$ [ns] | $I_3$ (%) | $\tau_1$ [ps] | $I_1$ (%) |
|---|---|---|---|---|---|---|
| 9.7 nm | 341 ± 4 | 89 ± 4 | 2.6 ± 0.2 | 3.4 ± 0.1 | 208 ± 10 | 7 ± 1 |
| 7.1 nm | 354 ± 3 | 95 ± 3 | 2.7 ± 0.3 | 1.5 ± 0.2 | 550 ± 50 | 4 ± 2 |
| 4.8 nm | 373 ± 3 | 98 ± 1 | 2.0 ± 0.1 | 1.4 ± 0.1 | - | - |
| 2.8 nm | 351 ± 4 | 95 ± 2 | 2.7 ± 0.2 | 4.4 ± 0.2 | - | - |

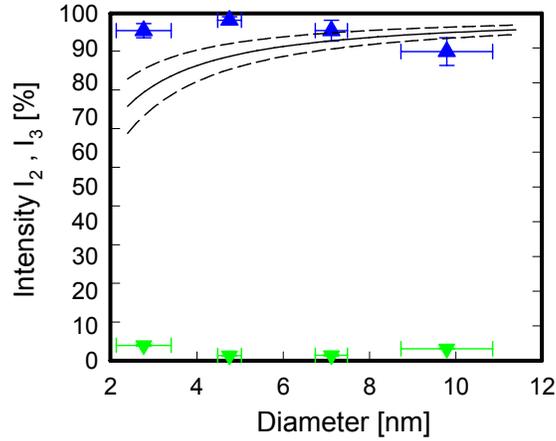

**Figure S6**. (a) Positron lifetime intensities $I_2$ (▲) and $I_3$ (▼) as a function of nanocrystal diameter. The curved lines denote estimated fractions of positrons stopped in the PbSe cores assuming interparticle distances of 1.5 nm (solid line), 1.1 nm (upper dashed line) and 1.9 nm (lower dashed line).

We estimated the relative fractions of positrons $f_{PbSe}$ stopped inside the PbSe nanoparticles and the fraction $f_{OA} = 1 - f_{PbSe}$ inside the oleic acid (OA) capping ligands as follows. Assuming a spherical shape of the PbSe core of the nanoparticles of diameter $d$ and a spherical shell of ligands of thickness $\Delta/2$ (with $\Delta$ the interparticle distance), we can estimate the fraction $f_{PbSe}$ of positrons stopped inside the PbSe core of the nanoparticles using: $f_{PbSe} = \dfrac{\rho_{PbSe} d^3}{(\rho_{PbSe} - \rho_{OA})d^3 + \rho_{OA}\Delta^3}$, where densities of

$\rho_{PbSe} = 8.1$ g/cm$^3$ and $\rho_{OA} = 0.9$ g/cm$^3$, and the measured interparticle distance of $\Delta = 1.5$ nm were used.[1] Note that the interparticle distance may reduce significantly (down to $\Delta = 1.1$ nm) upon removal of ligands by prolonged washing.[1]

### D. 2D-ACAR and Doppler broadening experiments

The nanocrystal films were studied with a high-intensity slow positron beam (POSH) originating from pair production in the reactor at Delft,[8] while the bulk PbSe single crystal was studied using a $^{22}$Na radioactive source, which emits a β-spectrum of positrons up to ~0.54 MeV. The 2D-ACAR measurements were carried out using an Anger-camera-type setup.[8,9] In order to select the optimum positron implantation energies for use in the 2D-ACAR experiments on the PbSe NC films, depth-profiling Doppler Broadening of Annihilation Radiation measurements were performed using the $^{22}$Na based Variable-Energy Positron (VEP) beam at positron implantation energies ranging from 0.1 keV to 25 keV. The Doppler depth-profiles were analyzed using the VEPFIT software package.[10]

### E. Computational Details

**E1. Positron-Electron Momentum Density (PEMD) of rock salt PbSe**

PbSe has the NaCl structure with space group $Fm\bar{3}m$ (No. 225) with lattice parameter $a = 0.6120$ nm.[11] Our electronic structure calculations are based on the local density approximation of density functional theory. The crystal potential was obtained by means of an all-electron fully charge-self-consistent semi-relativistic electronic structure computation using the Korringa-Kohn-Rostoker (KKR) method[12,13] with Von Barth-Hedin exchange and correlation.[14] The positron band structure and ground-state wave function were computed the same way, using the inverted electronic Hartree potential and adding a positron-electron correlation potential. The WIEN2K package[15] was employed to carry out calculations using the full-potential linear augmented plane wave (FLAPW) scheme in order to ascertain that the KKR band structure and Fermi level are basically the same as the full-potential results.

The positron-electron momentum density (PEMD), $\rho^{2\gamma}(\mathbf{p})$, is given in the independent particle model by

$$\rho^{2\gamma}(\mathbf{p}) = \pi r_0^2 c \sum_i n_i(\mathbf{k}) \left| \int e^{-i\mathbf{p}\cdot\mathbf{r}} \psi_i(\mathbf{r}) \varphi_+(\mathbf{r}) d\mathbf{r} \right|^2,$$

where $r_0$ is the classical electron radius and $c$ the speed of light. The occupation function $n_i(\mathbf{k})$ equals 1 if the electron state $\mathbf{k}_i$ is occupied and 0 when it is empty, $\psi_i(\mathbf{r})$ is the electronic wave function for the state $\mathbf{k}_i$, $\varphi_+(\mathbf{r})$ the wave function of the positron in its ground state, $\mathbf{p} = \mathbf{k}+\mathbf{G}$ with $\mathbf{G}$ a reciprocal lattice vector, and the summation extends over all occupied $\mathbf{k}$ states. Wave functions and momentum density were calculated on a three-dimensional momentum mesh with $\Delta p = 2\pi/16a = 0.03396$ a.u. along all three coordinate axes. The momentum density contains contributions from both semi-core states and the filled valence band. The positron is repelled by the nuclei and has a greater chance to annihilate with the relatively spread out valence electrons than with the electrons bound in the ionic cores. The KKR formalism for calculating $\rho^{2\gamma}(\mathbf{p})$ is discussed in Refs. 12 and 13 and references therein. To enable a comparison with experiment the computed momentum density was projected onto the (001) plane. The two computed momentum distributions have been convoluted with the experimental resolution of 0.15×0.14 a.u.² full-width-at half-maximum. A 1D-ACAR distribution is related to the 3D positron-electron momentum density $\rho^{2\gamma}(\mathbf{p})$ by two projections following: $N(p_z) = \iint \rho^{2\gamma}(\mathbf{p}) dp_y dp_x$ and is in practice obtained by projection of a calculated or measured 2D-ACAR distribution $N(p_x, p_z)$ using $N(p_z) = \int N(p_x, p_z) dp_x$.

Figure S7 presents the calculated 1D-ACAR momentum distribution for defect-free bulk PbSe. The 1D-ACAR profile for PbSe provides an insight into the origin of the peak at ~1 a.u. in the ratio curves shown in Fig. 2. Following Weber et al.[16], one can define a Fermi-like momentum of the 1D-ACAR profile near 1 a.u. (at the so-called Jones-zone boundary[16] for semiconductors) where the electrons have mostly Se(4p) character. The smearing out in momentum space and hence in the 1D-ACAR due to electron confinement leads to the behaviour observed near 1 a.u. in the ratio plot of Fig. 2.

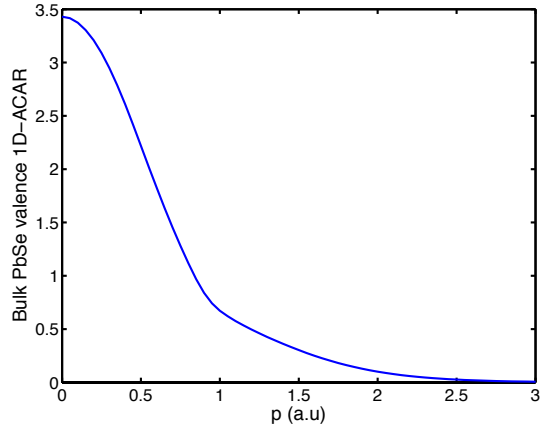

**Figure S7.** Calculated 1D-ACAR for defect-free bulk PbSe.

### E2. Calculated PEMD of atomic oxygen

Figure S8 shows the calculated positron-electron momentum distribution $N(p)$ of atomic oxygen presented as a ratio to the calculated directionally averaged 1D-ACAR distribution of single crystal PbSe. The ratio curve for the case of 50% annihilations with atomic oxygen and 50% annihilations with PbSe is also presented. The latter implies a O(1s) contribution – responsible for the high momentum rise beyond $p = 2$ a.u. – of about 1%, which is of similar magnitude as the O(1s) contribution found in Positron Auger Electron Spectroscopy (PAES) measurements of surface oxidized Cu.[17] Previous studies have shown that PAES is an extremely surface sensitive analysis technique.[18]

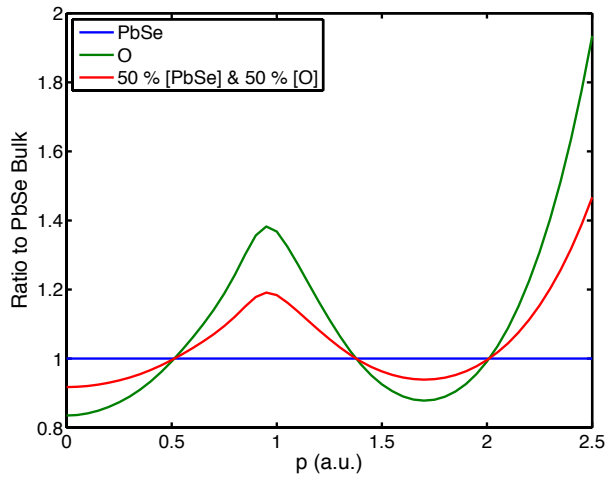

**Figure S8.** Calculated 1D-ACAR momentum distribution $N(p)$ of oxygen (O) presented as ratio to the calculated 1D-ACAR distribution of a bulk PbSe.

### E3. Calculation of the formation energy of the Schottky pair defect ($V_{Ps}V_{Se}$) in PbSe

In order to obtain an estimate for the concentration of divacancies, the formation energy of a Schottky cluster (adjacent Pb-Se vacancies) was calculated using the ab initio VASP code[19] (PAW-GGA-PBE potentials, 64-atom 2x2x2 supercell, 8x8x8 k-mesh, cut-off energies for the plane wave and augmentation functions of 450 and 750 eV, respectively). A formation energy of 1.47 eV was found after full relaxation, which implies that even at the highest synthesis temperature of 165 °C, the equilibrium concentration of Schottky pair defects is less than $10^{-16}$ per pair of Pb-Se atoms.

## F. Variation of the confinement peak with particle size

Figure S9 shows that the peak at $p\sim 1.0$ a.u. in Fig. 2 varies approximately as $1/d$ on particle diameter $d$, as could be expected from the scaling law for the optical band gap $\Delta E_{gap} \sim 1/d$ for PbSe nanoparticles.[20] The small increase of area for the 9.7 nm NCs is probably due to the other effect of the confinement of the positron wave function described by A. Calloni et al.[21].

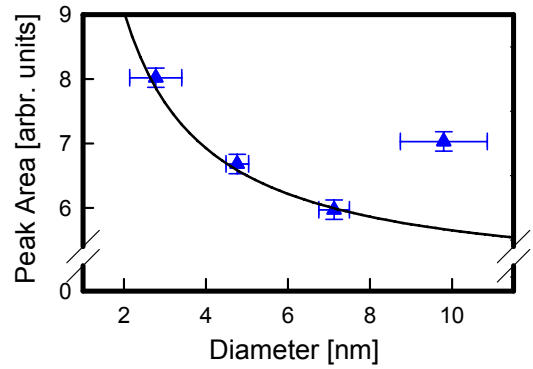

**Figure S9.** Area of the peak at $p\sim 1.0$ a.u. in Fig. 2 as a function of particle diameter. The solid curve shows a $1/d$ dependence for comparison.

## G. Correlation of Pb(5d) contribution to the PEMD and positron lifetime $<\tau>$

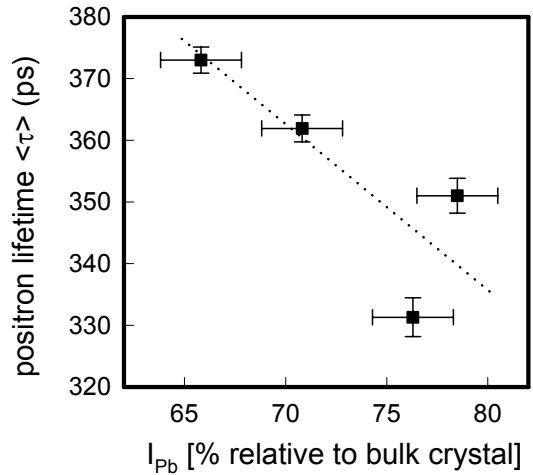

**Figure S10.** Average of the short positron lifetimes $<\tau>$ as a function of the Pb(5d) electron orbital contribution $I_{Pb}$.